\documentstyle[prl,aps,multicol,epsf,rotate]{revtex}
\begin{document}
\draft \title{Renormalized Field Theory of Polyelectrolyte Solutions}
\author{${\mbox{Michael Stapper} }^{}$ and 
${\mbox{Tanniemola B. Liverpool}}^{}$}
\address{
Max-Planck-Institut f\"ur Polymerforschung,
Postfach 3148, D-55021 Mainz, Germany
}
\date{\today}
\maketitle
\begin{abstract}
  We present a field-theoretic renormalization group (RG) analysis of a dilute
  solution of flexible, screened polyelectrolyte chains (Debye-H\"uckel
  chains) in a polar solvent. We perform a virial expansion to calculate the
  scaling dependence of the Osmotic pressure on the chain density, chain
  size and screening length. The formalism is developed for general chain
  distributions and we give explicit forms for a monodisperse chain size 
  distribution. We utilize the fact that a Debye-H\"uckel chain may be mapped
  to a {\it local} field theory with {\bf two} length-scales {\it requiring}
  the calculation of scaling functions {\it as well} as exponents to fully
  describe its scaling behaviour.
\end{abstract}
\begin{multicols}{2}
  Polyelectrolytes are polymers with ionizable monomers which in polar
  solvents, such as water, dissociate into charged polyions and small
  `counter-ions' of opposite sign. They are of widespread importance with many
  applications in physics, biology and chemistry~\cite{Joanny}. Their
  solublility in polar solvents means that they also have many industrial
  applications (e.g as super-absorbers). Typical examples are DNA and
  sulphonated polystyrene.
  
  Polyelectrolytes are quite poorly understood, mainly because it has been
  difficult to deal with their long-range coulomb interactions theoretically.
  In addition, screening, complex-formation, counter-ion condensation and salt
  and the comparable length-scales of their interactions make comparison
  between theory and experiment difficult.  Neutral polymers on the other
  hand, have been extremely succesfully described by scaling ideas and
  renormalization group theories~\cite{frdec}.  We recently proposed a new
  approach to the renormalization group applied to the study of {\it weakly}
  charged {\it flexible} Debye-H\"uckel (DH) polyelectrolytes in
  solution~\cite{useuro}.
  
  We first review some of the results and the philosophy of the previous
  paper. We found that the `range' of the interaction, the Debye screening
  length, was an important new length-scale {\it independent} of the system
  size.  This is fundamentally different to the systems usually studied in
  conventional statistical mechanics with short-ranged interactions where the
  size of the system or volume is the only relevant macroscopic (low $q$
  cut-off) length-scale and all extensive quantities scale with this volume.
  We obtained non-trivial scaling of quantities like the end-to-end distance
  and radius of gyration of isolated chains with the number of monomers and
  with the screening length~\cite{useuro,mickkrem}. This second scale also
  means that naive scaling arguments are liable to break down.  These
  complications are not new: it has long been understood ~\cite{oosawa} that
  taking the thermodynamic limit is problematic in systems with long-range
  Coulomb interactions.
  
  In the literature regarding flexible chains with unscreened coulomb
  interaction, it is generally agreed that the chains are in a `rigid' regime
  in $d=3$.  It is then tempting to suggest that even if one has screened
  interactions there is a length-scale much below that of the screening length
  where the interaction is effectively unscreened and thus the chain would be
  in a rigid state and not amenable to description by a flexible chain model.
  An important point has been overlooked in this analysis: the physical
  arguments leading to the rigid regime depend on the region of the chain
  being considered being `extensive' i.e. of the order of the system size, but
  if one is dealing with screened interactions the region of the chain where
  the interaction is `effectively' unscreened is {\it by definition} not
  extensive.  There are fluctuations happening on much {\it longer}
  length-scales.  In experiments polyelectrolytes are found to be extended but
  {\it not} rigid~\cite{Joanny}. We conclude that there is, in principle, no
  problem with a flexible model being used to describe realistic screened
  polyelectrolytes as long as one is studying the long-wavelength behaviour.
  
  Another critique of flexible chain models of polyelectrolytes is related to
  the phenomenon of `electrostatic rigidity'. This is due to the repulsion of
  regions of the chain far from each other. Chains which are {\it much longer}
  than the screening-length can be partitioned into flexible coils or `Pincus
  blobs' which align into a rod-like blob-pole~\cite{Joanny}.  Such long
  chains are outside the regime of our theory. In fact it has recently been
  shown that the number of monomers of polyelectrolyte required for the
  polyion to be long enough to be in this rigid Pincus blob-pole in normal
  solvents is several orders of magnitude larger than any physically or
  computationally realisable polyelectrolytes~\cite{mickkrem,nyrkova}. The
  indication seems from experiment {\it and} simulations that the viable and
  interesting region of polyelectrolyte behaviour is that when the
  screening-length and the chain size are comparable and {\it large}
  (macroscopic) - exactly that region being addressed by our theory. The first
  letter addressed single chain properties.  In this Letter, we apply this
  formalism to the calculation of contribution of the polyions to the osmotic
  pressure at non-zero chain concentrations, a problem of experimental
  relevance.
%We perform a virial expansion thus restricting ourselves to dilute
%solutions. An extension to the semi-dilute case will be presented
%elsewhere~\cite{longer}. 

  In the presence of counter-ions and/or salt the coulomb interaction becomes
  screened. The simplest and most widely used model of a screened
  polyelectrolyte which we use as the starting point for our analysis is a
  solution of flexible chains in $d$ dimensions with a Debye-H\"uckel (DH) or
  Yukawa interaction between charged monomers~\cite{useuro}. This is given by
  $V({\bf r})= g\int \frac{d^{d}q}{(2\pi)^{d}} \frac{e^{i{\bf q \cdot
        r}}}{q^{2}+\kappa^{2}}$ which has the familiar form $V({\bf r})=
  g{e^{-\kappa|{\bf r}|}}/{|{\bf r}|}$ in $d=3$. The chains are described by a
  set of vectors $\{{\bf r}_{j}(s)\}$ parametrised by the arc-length $s$ and
  label $j$.  We define a `{\it Kuhn-segment}' length of the chain, $\ell$ and
  $L_{j}= \ell^{2} N_{j}$ is the size (area) of the $j$th chain, where $N_{j}$
  is the number of segments on that chain. The Kuhn-length here refers to the
  distance between {\it charged} monomers. In $d=3$ the coupling constant $g
  \propto \lambda_{B}=\beta\frac{q^{2}}{4\pi\varepsilon\varepsilon_{0}}$ is
  proportional to the Bjerrum length, $\lambda_{B}$ which measures the
  strength of the bare coulomb interaction, where $1/\beta=k_{B}T$. In water,
  $\lambda_{B}=7.14 \mbox{\AA}$. As usual $\varepsilon\varepsilon_{0}$ is the
  dielectric constant of the solvent, $k_{B}$ is the Boltzmann constant and
  $T$ the temperature. The screening length or range of the interaction is
  given by $1/\kappa$ and $\kappa$ is a function of the density of screening
  ions and the dielectric properties of the solution.  The flexible chain DH
  model is thought to be valid only when the polyelectrolyte is weakly charged
  and in a non-zero ion density~\cite{Vilgis,Muthukumar,liu}.
  
  We calculate the contribution of the polyions to the osmotic pressure for a
  solution of chains of $N$ monomers and screening length $1/\kappa$ and
  chain density $\rho_{p}$. We find the general scaling form
\begin{eqnarray}
\frac{\Pi_{p}(\rho_{p},N,\bar{\kappa})}{\rho_{p}k_{B}T} &=& 1 +
F_{1}(\rho_{p}\ell^{d}/\bar{\kappa}^{d},N\bar{\kappa}^{1/\nu_{\phi}}) \nonumber \\
&=& 1 + F_{2}(\rho_{p}\ell^{d}N^{d\nu_{\phi}},\bar{\kappa}^{2}N^{2\nu_{\phi}}).
\label{genscal}\end{eqnarray}
This scaling law is the screened polyelectrolyte equivalent of the des
Cloizeaux scaling law~\cite{ftsoln} for neutral polymers.
From the virial expansion we find explicitly that for a monodisperse
distribution of chains of $N$ monomers
\begin{equation} \frac{\beta \Pi_{p}}{\rho_{p}} =  1 + \frac{3D
    \rho_{p}^{}\ell^{d}}{16\bar{\kappa}^{d}} (\epsilon N^{2}\bar{\kappa}^{2/\nu_{\phi}}+  \epsilon^{2} P(N\bar{\kappa}^{1/\nu_{\phi}})) +
  O(\rho_{p}^{2}),\label{osmotic}\end{equation}
where $\epsilon=6-d$, $\nu_{\phi} = 2/(d-2)$ and $D$ is a non-universal
constant.  $P(x)$ is a complicated function and we have defined a
dimensionless screening variable measuring the screening-length in terms of
the Kuhn-length, $\bar{\kappa} \equiv \kappa\ell$.  These are the main results
of our letter.  What follows is a brief outline of the calculation; details
will be presented in a longer future publication~\cite{longer}.

We find it convenient to work in the Grand Canonical Ensemble of chains with
chemical potentials controlling the monomer and chain fugacities. The grand
canonical partition function is given by
\end{multicols}
\begin{equation}
Z[\{\sigma\}] = 1 + \sum_{p=1}^{\infty}\frac{1}{p!}\int^{\infty}_{0}dL_{1} \ldots dL_{p}
e^{\sigma(L_{1})} \ldots e^{\sigma(L_{p})}\bar{Z}[L_{1},\ldots,L_{p},\kappa]
\end{equation}
where $\sigma(L)$ is the fugacity of a polyion of size (and not necessarily a
linear function of) $L$ and
\begin{equation} \bar{Z}[L_{1},\ldots,L_{p},\kappa] = \int
  \prod_{j=1}^{p}{\cal D}[{\bf r}_{j}(s)]e^{- \frac{1}{2}\sum_{j=1}^{p}
    \int^{L_{j}}_{0} ds \left({\partial {\bf r}_{j}}/{\partial s}\right)^{2}-
    g \sum_{i,j=1}^{p}\int^{L_{i}}_{0}\int^{L_{j}}_{0} ds ds' V({\bf
      r}_{i}(s)-{\bf r}_{j}(s'))}.\end{equation}

Following de Gennes~\cite{deGennes72} and des Cloizeaux~\cite{ftsoln} and
performing an Hubbard-Stratonovich transformation~\cite{Pfeuty,Jug}, this may
be mapped to a local-field theory with {\it two} sets of fields with a
`magnetic' field coupled to one set of fields. The grand canonical partition
function, ${Z}[\{\sigma\}]$ is given by
\begin{equation}
{Z}[\{\sigma\}]= {\cal N}\int \prod_{j=1}^{}[{\cal
  D}\phi_{j}][{\cal D}\psi]e^{-\int_{\bf r}\left[ \frac{1}{2}\sum_{j=1}^{} \left\{
    (\nabla \phi_{j})^{2}+ t_{j}\phi_{j}^{2}  + u \phi_{j}^{2}\psi \right\}+\frac{1}{2}(\nabla
  \psi)^{2}+\frac{1}{2} \kappa^{2} \psi^{2}
+ \sum_{j=1}^{} H_{j} \cdot \phi_{j}^{1} \right]},  \label{action}
\end{equation} 
\begin{multicols}{2}
  where the $\phi_{j}({\bf r})$ are $n \rightarrow 0$ component (polyion)
  fields with $O(n)$ symmetry and $\psi({\bf r})$ is a scalar (coulomb) field.
  The $t_{j}$ are the polyion monomer fugacities, $1/\kappa^{2}$ is the
  screening length and ${\cal N}=
  (2\pi)^{-1/2}(\mbox{det}\{\nabla^{2}+\kappa^{2}\})^{1/2}$ is a normalisation
  factor.  The coupling is given by $u=2 i\sqrt{g}$. The resulting field
  theory is different to that from neutral polymer solutions~\cite{frdec}
  where we have only one type of field and the interaction potential is
  quartic $\int_{r}(\phi_{j}^{2})^{2}$. Here we have {\it two} types of field
  and a cubic interaction $\int_{r}\phi_{j}^{2}\psi$~\cite{Janssen}. The upper
  critical dimension $d_{c}=6$ follows from a dimensional analysis of the
  action in equation (\ref{action}).  The `magnetic' field $H_{j}$ which
  without loss of generality we align along the first component $\phi_{j}^{1}$
  of the polyion fields, is related to the polyion fugacity, $ e^{\sigma(L)}
  \equiv f(L) = \sum_{i=1}^{} e^{- L t_{i}} \frac{1}{2} H_{i}^{2}$.  It is
  useful to rewrite this in the continuum form $f(L) = \int \frac{dt}{2 \pi i}
  e^{- L t} \frac{1}{2} h^{2}(t)$ making the association $\frac{\Delta t}{2
    \pi i}h^{2}(t_{j})=H^{2}_{j}$.
  
  We want to calculate the contribution of the polyions to the osmotic
  pressure, $\Pi_{p}$ given by
\begin{equation}
\Pi_{p} = k_{B}T \log Z[\{\sigma\}].
\end{equation}
There is a simple relation between the polyion fugacity $\sigma(L)$, the
monomer density $\rho_{p}$ and the chain-size distribution. The number density
of of polyions of size $L$ is given by
\begin{equation}
C_{p}(L)= \frac{\delta \ln Z[{\sigma}]}{\delta \sigma(L)} =
\sum_{j}f(L)\frac{\delta H_{j}^{2}}{\delta f(L)}\frac{\delta \ln Z[{f}]}{\delta H_{j}^{2}}\label{distrib}
\end{equation}
For a monodisperse solution of polyions of size $L_{0}$, 
\begin{equation} C_{p}(L)= \rho_{p} \delta(L-L_{0}).
\label{monodisp}\end{equation}
We can easily generalise our results to arbitrary distributions by changing the form of $C_{p}(L)$.
We perform a virial expansion of the osmotic pressure calculating the first
and second virial coefficients using renormalized perturbation theory. This
entails doing a double expansion in $H$ and $u$.   
A typical term in our perturbation series to $m$-th order $H$
and $l$-th order in $u$ (with $m,l$ even) is given by
\end{multicols}
\begin{eqnarray}
&& u^{l}\frac{1}{m!}\frac{1}{2^{l} l!}\sum_{\mbox{all pairs}\{j_{1}...j_{m}\}}H_{j_{1}} \ldots
H_{j_{m}}\left\langle \phi_{j_{1}}^{1}(0) \ldots
  \phi_{j_{m}}^{1}(0)\left(\int_{{\bf k,q}}\phi_{i}({\bf -k-q})\cdot \phi_{i}({\bf k})\psi({\bf q})\right)^{l}\right\rangle_{H=0}^{C} \nonumber 
\end{eqnarray}
\begin{multicols}{2} where $C$ refers to connected diagrams. The series for
  $\Pi_{p}$ can be calculated using standard diagrammatic methods and we show
  the diagrams required to calculate the first and second virial coefficients
  in Fig. 1.
  
  Renormalization proceeds via the absorption of ultra-violet divergences in
  Z-factors which are sums of poles in $\epsilon=6-d$.  We use a dimensional
  regularisation scheme with minimal subtraction of poles~\cite{ft}.  We can
  express the renormalized parameters in terms of the bare parameters
\begin{equation} \begin{array}{cl}
u & = \mu^{\epsilon/2}S_{d}^{-1}Z_{u}u^{R}\\
\kappa^{2} & = \mu^{2}(\kappa^{2})^{R} \\ 
t_{j} & = \mu^{2}{Z_{\phi}^{-1}} \left[
  t_{j}^{R} Z_{\phi,\phi^{2}}+(\kappa^{2})^{R} Z_{\phi,\psi^{2}}\right] \\
H_{j} & = \mu^{\frac{d+2}{2}}{Z_{\phi}^{-1/2}}H_{j}^{R},
\end{array}\end{equation}
where $\mu^{-1}$ is an external length-scale and $S_{d}=\Omega_{d}/(2\pi)^{d}$
with $\Omega_{d}$ the angular part of a d-dimensional volume integral.  It is
important to note the matrix renormalization of the polyion fugacities.  
%Sincethere are two independent fields, we have two sets of exponents
%$\nu_{\psi},\nu_{\phi};\eta_{\psi},\eta_{\phi}$. To distinguish between the
%two we use a superscript $\psi$ for the Coulomb terms and a superscript
%$\phi$ for the polyion terms. 
Because the chains renormalize independently, the RG flow and fixed point
values of $u^{R}*$ are the same as for the single chain~\cite{useuro}. In
particular, working at the non-trivial D-H fixed point, we calculate the
exponent $\nu_{\phi}=2/(d-2)$.

We define the renormalized osmotic pressure $\Pi_{p}/k_{B}T =
\mu^{d}\Pi_{p}^{R}\left(\rho_{p}^{R},L^{R},(\kappa^{2})^{R},u^{R}\right)$.  We
can solve it's RG equation to give us the general scaling form of $\Pi_{p}$.
We extract the scaling behaviour of the osmotic pressure by writing
$\Pi_{p}^{R}$ in terms of flowing variables, fixing the RG flow parameters by
a matching condition and then transforming back to the bare quantities. The
result is equation (\ref{genscal}).
The renormalized osmotic pressure can also be calculated to 1-loop order in a
virial expansion (see Fig. 1) as
\end{multicols}
\begin{equation} {\Pi_{p}^{R}} = \frac{1}{2}\int \frac{dt^{R}}{2 \pi i}
  \frac{(h^{R}(t^{R}))^{2}}{2}a_{1}(t^{R},\kappa^{R}) + \frac{1}{4!}\int
  \frac{dt^{R}_{1}}{2 \pi i}\int \frac{dt^{R}_{2}}{2 \pi i} \frac{(h^{R}(t^{R}_{1}))^{2}}{2} \frac{(h^{R}(t^{R}_{2}))^{2}}{2}
a_{2}(t^{R}_{1},t^{R}_{2},\kappa^{R}) + O((h^{R})^{6}),
\label{virial}\end{equation} 
  where $a_{1}(t,\kappa)$ is the contribution from the 2-point diagrams and
  $a_{2}(t_{1},t_{2},\kappa)$ is the contribution from four-point diagrams.
  Using equations (\ref{distrib}), (\ref{monodisp}) and (\ref{virial}),
  performing multiple inverse Laplace transforms, we find for a monodisperse
  solution of polyions of size $L$, 
\begin{equation} {\Pi_{p}^{R}} =
  \rho_{p}^{R}\left[ 1 - \frac{1}{24}\rho_{p}^{R}\bar{A_{1}}^{-2}(L^{R},\kappa^{R})\bar{A_{2}}(L^{R},
    L^{R}, \kappa^{R}) + O((\rho_{p}^{R})^{2}) \right] \end{equation} where
  $\bar{A_{1}}(L,\kappa)= \int \frac{dt}{2 \pi i} e^{tL} a_{1}(t,\kappa)$ and
  $\bar{A_{2}}(L_{1},L_{2},\kappa)= \int \frac{dt_{1}}{2 \pi i}
  \frac{dt_{2}}{2 \pi i} e^{t_{1}L_{1}}e^{t_{2}L_{2}}
  a_{2}(t_{1},t_{2},\kappa)$. 
  We find explicitly that $\bar{A_{1}}^{-2}(L,\kappa) =
  A_{1}^{-2}(L\kappa^{1/\nu_{\phi}})= \frac{1}{4}\left[1 + 3 \epsilon e^{L
      \kappa^{1/\nu_{\phi}}} E_{1}(L\kappa^{1/\nu_{\phi}})\right] $ and $
  \bar{A_{2}}(L,L,\kappa) = A_{2}(L\kappa^{1/\nu_{\phi}})= - 18 \epsilon
  (L\kappa^{1/\nu_{\phi}})^{2} - 27 \epsilon^{2} \left[ (e^{L
      \kappa^{1/\nu_{\phi}}}E_{1}(L\kappa^{1/\nu_{\phi}}) +
    1/4)\left(1+(L\kappa^{1/\nu_{\phi}})^{2}\right) \right] - 27 \epsilon^{2}
  \left[(L\kappa^{1/\nu_{\phi}}+ 1)(\ln(L\kappa^{1/\nu_{\phi}}) +\gamma) -
    \frac{1}{2}L\kappa^{1/\nu_{\phi}} - \frac{1}{4} e^{2
      L\kappa^{1/\nu_{\phi}}}\right]. $
\begin{multicols}{2} 
  Taking into account dimensional factors we obtain equation (\ref{osmotic})
  with $P(x)$ is the $O(\epsilon^{2})$ part of $A_{1}^{-2}(x)A_{2}(x)$.  It is
  reassuring that the explicit form satisfies the general solution in equation
  (\ref{genscal}).

We compare our result, equations (\ref{genscal},\ref{osmotic}) with the
simulations performed by Stevens and Kremer in $d=3$ on DH
polyelectrolytes~\cite{stevkrem}. They calculate the scaling behaviour of the
polyion contribution to the osmotic pressure with the monomer density
$\rho_{m} = \rho_{p} N$. Their data which is for salt-free solutions, i.e.
from charge neutrality $\kappa=(4\pi\lambda_{B}\rho_{m})^{1/2}$, has the
osmotic pressure as independent of chain length for both dilute and
semi-dilute monomer concentrations. For small concentrations they find
$\beta\Pi_{p} \sim \rho_{m}$ whilst for large concentrations they observe the
neutral polymer scaling $\beta\Pi_{p} \sim \rho_{m}^{9/4}$. This is consistent
with our general scaling law in equation (\ref{genscal}) evaluated at $d=3$
with $F_{1}$ given by $F_{1}(x,y) = -1+ f_{1} x^{}y^{2} + f_{2}x^{-4}y^{-3} +
g(x,y)$ where $f_{1}$ and $f_{2}$ are non-universal constants. The $f_{1}$
term (which is the source of the small $\rho_{m}$ behaviour) has exactly the
form as that in the leading term of our virial expansion, equation
(\ref{osmotic}). The term $g(x,y)$ measures the $N$ dependent cross-over
between the two regimes. We point out the interesting fact that from the
structure of the field theory, there are three fixed points governed by the
ratio $c = \ell^{2}{\kappa}^{2}N$~\cite{useuro}.  Because $\kappa \sim
\rho_{m}^{1/2}$, by varying $\rho_{m}$, we cross-over from the polyelectrolyte
($c \sim 1$) to the SAW fixed point ($c>>1$). Our calculation {\it explicitly}
captures this cross-over. The third fixed point (rod) occurs only at
$\rho_{m}=0$. One may wonder why although we have performed an expansion in
$\epsilon=6-d$ and $\epsilon=3$ in 3 dimensions, we get rather good agreement
with the simulation data. We suggest that this is due to the symmetry identity
of the vertex functions of the field theory which leads to the exact
calculation of the exponent $\nu_{\phi}$~\cite{Jug}. Our scaling law can also
be used for polyelectrolytes in a salt solution but there the leading
behaviour of $\kappa$ is {\it not} related to the monomer
density~\cite{longer}.

It is also interesting to analyse the form of the second virial coefficient
from equation (\ref{osmotic}). In the theoretical work on neutral polymers,
this quantity is used to define the `overlap' variable which usually has the
form (in $d=3$) $\langle R^{2}\rangle ^{3/2} \sim N^{3\nu_{1}}$ where $\langle
R^{2}\rangle$ is the mean end-to-end distance and $\nu_{1}=0.588$ is the size
scaling exponent of the Edwards SAW. Here we find that the overlap variable is
rather complicated function of the screening length, $1/\kappa$ and the chain
size $N$. This is reasonable for a system with long-range interactions as the
`effective' size of the particles strongly depends on the range of the
interaction as well as the physical size of the object. We see that the
overlap variable can in principle be somewhat larger than the chain size. This
is also consistent with the observations in simulations~\cite{stevkrem}.

  In summary, we have calculated the scaling behaviour of the contribution of
  the polyions to the osmotic pressure using field theoretic methods. We
  obtained a general scaling law of the osmotic pressure. To our knowledge,
  this is the first systematic attempt to calculate the osmotic pressure of a
  solution of Debye-H\"uckel chains without uncontrolled
  approximations~\cite{Muthukumar}. We also found that we can explain all
  the main features of the osmotic pressure measured in some recent
  simulations of screened polyelectrolytes.  At a more fundamental level, this
  calculation explicitly shows the subtleties of performing field theory with
  two {\it relevant} and {\it independent} length-scales and provides another
  interesting application of field theoretic methods to soft condensed matter
  systems.
  
  We acknowledge helpful discussions with R. Golestanian, H-K. Janssen, J-F.
  Joanny, K. Kremer, K. M\"uller-Nedebock, S. Obhukov, M. Rubinstein, M.
  Schmidt, M. Stevens and T. Vilgis.  We particularly thank K.  Kremer and T.
  Vilgis for a critical reading of the manuscript.
%

%

%%%%%%%%%%%%%%%%%%% FIGURE CAPTIONS %%%%%%%%%%%%%%%%%%
%%%%%%%%%%%%%%%%%%%%%%%%%%%%%%%%%%%%%%%%%%%%%%%%%%%%%%

%  
\begin{figure}
  \epsfxsize 8cm {\epsffile{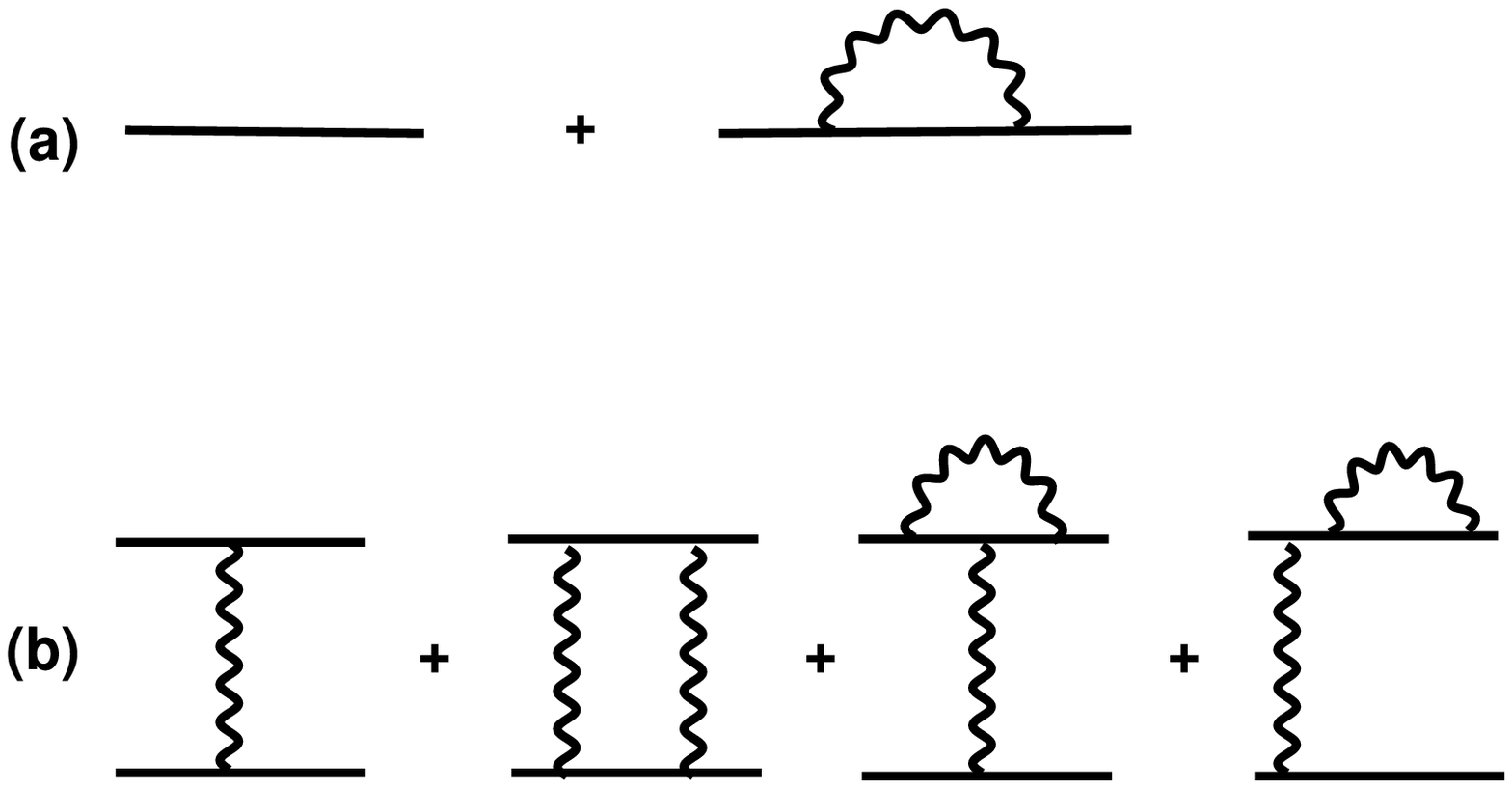}}
  
  \protect\begin{minipage}[t]{8.5cm}{\centerline{\caption{The diagrams for (a)
          $a_{1}(t,\kappa)$ and (b) $a_{2}(t_{1},t_{2},\kappa)$.  A straight
          line corresponds to a free polyion propagator $(q^{2}+t_{j})^{-1}$,
          a wavy line a coulomb propagator $(q^{2}+\kappa^{2})^{-1}$ and a
          vertex to $u/2$. We integrate momentum over all closed loops}}}
  \end{minipage} \label{fig1} \end{figure} 

\end{multicols}
\end{document}